\documentclass[
 aip,
 amsmath,amssymb,
preprint
]{./revtex4-2}
\usepackage{hyperref}
\usepackage{natbib}
\usepackage{graphicx}
\usepackage{dcolumn}
\usepackage{bm}
\usepackage{multirow}
\usepackage{amsmath}
\usepackage{makecell}
\usepackage{mathrsfs}
\usepackage{tikz}

\begin{document}

\preprint{PoF-submitted/manuscript}

\title{Optimizing Flow Control with Deep Reinforcement Learning: Plasma Actuator Placement around a Square Cylinder}

\author{Mustafa Z. Yousif}
\email[]{These authors contributed equally to this work.}
\affiliation{School of Mechanical Engineering, Pusan National University, 2, Busandaehak-ro 63beon-gil, Geumjeong-gu, Busan, 46241, Republic of Korea}

\author{Kolesova Paraskovia}
\email[]{These authors contributed equally to this work.}
\affiliation{School of Mechanical Engineering, Pusan National University, 2, Busandaehak-ro 63beon-gil, Geumjeong-gu, Busan, 46241, Republic of Korea}

\author{Yifang Yang}
\affiliation{School of Mechanical Engineering, Pusan National University, 2, Busandaehak-ro 63beon-gil, Geumjeong-gu, Busan, 46241, Republic of Korea}

\author{Meng Zhang}
\affiliation{School of Mechanical Engineering, Pusan National University, 2, Busandaehak-ro 63beon-gil, Geumjeong-gu, Busan, 46241, Republic of Korea}

\author{Linqi Yu}
\affiliation{School of Mechanical Engineering, Pusan National University, 2, Busandaehak-ro 63beon-gil, Geumjeong-gu, Busan, 46241, Republic of Korea}

\author{Jean Rabault}
\affiliation{IT Department, Norwegian Meteorological Institute, Postboks 43, 0313 Oslo, Norway}

\author{Ricardo Vinuesa}
\affiliation{FLOW, Engineering Mechanics, KTH Royal Institute of Technology, Stockholm, 10044, Sweden}

\author{Hee-Chang Lim}
\email[]{Corresponding author, hclim@pusan.ac.kr}
\thanks{}
\affiliation{School of Mechanical Engineering, Pusan National University, 2, Busandaehak-ro 63beon-gil, Geumjeong-gu, Busan, 46241, Republic of Korea}

\date{\today}

\begin{abstract}
The present study proposes an active flow control (AFC) approach based on deep reinforcement learning (DRL) to optimize the performance of multiple plasma actuators on a square cylinder. The investigation aims to modify the control inputs of the plasma actuators to reduce the drag and lift forces affecting the cylinder while maintaining a stable flow regime. The environment of the proposed model is represented by a two-dimensional direct numerical simulation (DNS) of a flow past a square cylinder. The control strategy is based on the regulation of the supplied alternating current (AC) voltage at three distinct configurations of the plasma actuators. The effectiveness of the designed strategy is first investigated for Reynolds number, $Re_{D} = 100$, and further applied for $Re_{D} = 180$. The applied active flow control strategy is able to reduce the mean drag coefficient by 97\%  at $Re_{D} = 100$ and by 99\%  at $Re_D=180$. Furthermore, the results from this study show that with the increase in Reynolds number, it becomes more challenging to eliminate vortex shedding with plasma actuators located only on the rear surface of the cylinder. Nevertheless, the proposed control scheme is able to completely suppress it with an optimized configuration of the plasma actuators.  
\end{abstract}

\maketitle

\section{Introduction}\label{sec:introduction}
The control of fluid flows is a critical problem in many industrial and engineering applications, including aerospace, automotive, and energy systems. Flow manipulation can serve various objectives, such as separation delay, transition delay, drag reduction, lift enhancement etc.\cite{GadeIHak1996} Particularly in the past few decades, flow control strategies around bluff bodies, such as circular and square cylinders, have been extensively studied, focusing on drag and lift reduction. \cite{Saha2003, Marquet2008, Feng2010} Although most of the studies have been done on circular cylinders, square cylinders are more commonly encountered in practical engineering applications, such as buildings, bridges, and offshore structures. The behavior of the flow past a square cylinder at low Reynolds numbers is steady and symmetric, with fluid particles moving uniformly and forming symmetrical vortices in the wake region. However, as the Reynolds number increases and reaches its first critical value, typically around $Re_{D} = 47$, the flow becomes unstable and asymmetric, undergoing vortex shedding. This shedding phenomenon leads to significant drag and lift forces acting on the cylinder, which can have adverse effects.\cite{Zhu2021} Hence, managing and reducing these forces through flow control methods is of utmost importance in order to optimize the performance of square cylinders in various applications.\par

The control strategies applied for fluid flow problems can be broadly categorized into passive and active flow control (AFC). Passive flow control methods use fixed geometry modifications to influence fluid flow around an object that can be efficient in terms of energy consumption and safety. However, they are limited in their effectiveness and need to be adaptable to changes in flow conditions. Among them, several studies have been done including the use of a splitter plate on the wake flow, usage of the micro vortex generators to control the transition to turbulence, and drag reduction for various spanwise modulations of the trailing edge. \cite{Bearman1965, Celik2008, Fransson2006, Tanner1972} In contrast, AFC methods involve actuators that require additional energy input to manipulate the flow of fluids in real time. They are mainly divided into active open-loop and active closed-loop actuation. Open-loop control involves pre-programmed actuation patterns, while closed-loop involves feedback control, where actuators are controlled based on feedback signals from sensors monitoring the flow field. \cite{Pastoor2008, Erdmann2011} These flow control approaches have been explored for decades. The most frequently employed strategies in AFC methods involve manipulating the fluid motion through techniques such as blowing and suction, as well as employing synthetic jet and co-flow jet-based methods. \cite{Arcas2004, Glezer2002, Liu2016} However, the implementation of these kinds of methods can add more complexity and cost to the system, especially for large-scale applications. On the other hand, another attractive approach that proved to be efficient for active flow control is based on dielectric barrier discharge (DBD) plasma actuators. \cite{Corke2010} They have been successfully used in various flow problems, such as flow separation control, turbulence suppression, drag reduction, and lift enhancement. \cite{Anzai2017, Mahfoze2017, Huang2021, Sato2019} The previous studies investigated the effects of different parameters, such as plasma actuator placement, frequency, the amplitude of applied alternating current (AC) voltage, and its waveform on the flow control performance. Additionally, plasma actuators are relatively easy to implement, pointing they can be incorporated into existing systems without requiring significant modifications. \par

Nonetheless, due to the complicity of the flow behavior represented by the high non-linearity of the Navier–Stokes equations, developing sophisticated control schemes that consider such a highly nonlinear problem is a challenging topic. With the rapid development of machine learning algorithms and computational power, promising results have been achieved by applying machine learning in various fields, such as image processing, natural language processing, robotics, and weather forecasting. Its capabilities for approaching highly nonlinear issues also found an application in several problems in fluid mechanics. As an example, impressive results have been achieved by employing machine learning in complex problems, such as reduced-order modeling, prediction of turbulent flows, turbulence modeling, and super-resolution reconstruction of turbulent flows.\cite{Kutz2017, Duraisamy2019, Lee2019, Brunton2020, Yousif2021_1, Kim2021, Yousif2022_2, Yousif2022_3, Yousif2022_4, Yousif2022_5} Furthermore, researchers proposed a deep reinforcement learning (DRL) approach for controlling complex systems with high-dimensional state and action spaces, which are commonly found in robotics, autonomous vehicles, and other real-world applications. DRL-based approaches aim to use deep neural networks to learn an optimal control policy through trial-and-error interactions with the designed environment. For instance, this models have been applied for a flow control problem, exhibiting promising outcomes in terms of effective control of two-dimensional cylinder flow employing two small jets, control of confined cylinder wakes, synthetic jet control applied to flow over an airfoil, and control of the turbulent fluid flows enclosed in a channel.\cite{Rabault2019_1, Rabault2019_2, Han2020, Tang2020, Li2021, Guastoni2023} Common reinforcement learning algorithms utilized for control problems, as well as their recent advances are discussed in a work of C. Vignon $et al.$\cite{Vignon2023} These studies have encouraged further investigation of DRL-based models for utilizing diverse control devices and various flow configurations. \par

In this paper, we investigate the application of the DRL-based approach in AFC of the flow past a square cylinder using multiple plasma actuators. Due to the high non-linearity associated with the proposed control configuration, which involves multiple plasma actuators operating within a specific range of applied AC voltage for each actuator, a neural network is trained to learn an optimal control policy, where the control strategy is based on the regulation of the supplied AC voltage amplitude at three distinct configurations of the plasma actuators. \par

This paper is organized as follows: In Section 2, we introduce the proposed DRL-based AFC model, which consists of three main components:the modeling of the incorporated plasma actuators, the numerical setup of the simulation environment, and the DRL-based training framework. In Section 3, we investigate the instantaneous and statistical results at two Reynolds numbers and analyze the model using the SHapley Additive exPlanations (SHAP) values method\cite{Lundberg2017} to gain insight into the effectiveness of the distributed probes in the flow field. Finally, in Section 4, we summarize the conclusions of this study. \par

\section{METHODOLOGY}\label{sec:Methodology}
\subsection{Modeling of plasma actuators}\label{subsec:II2}

\begin{figure}
\centering 
\scalebox{0.8}{\includegraphics[angle=0, trim=0 0 0 0, width=0.8\textwidth]{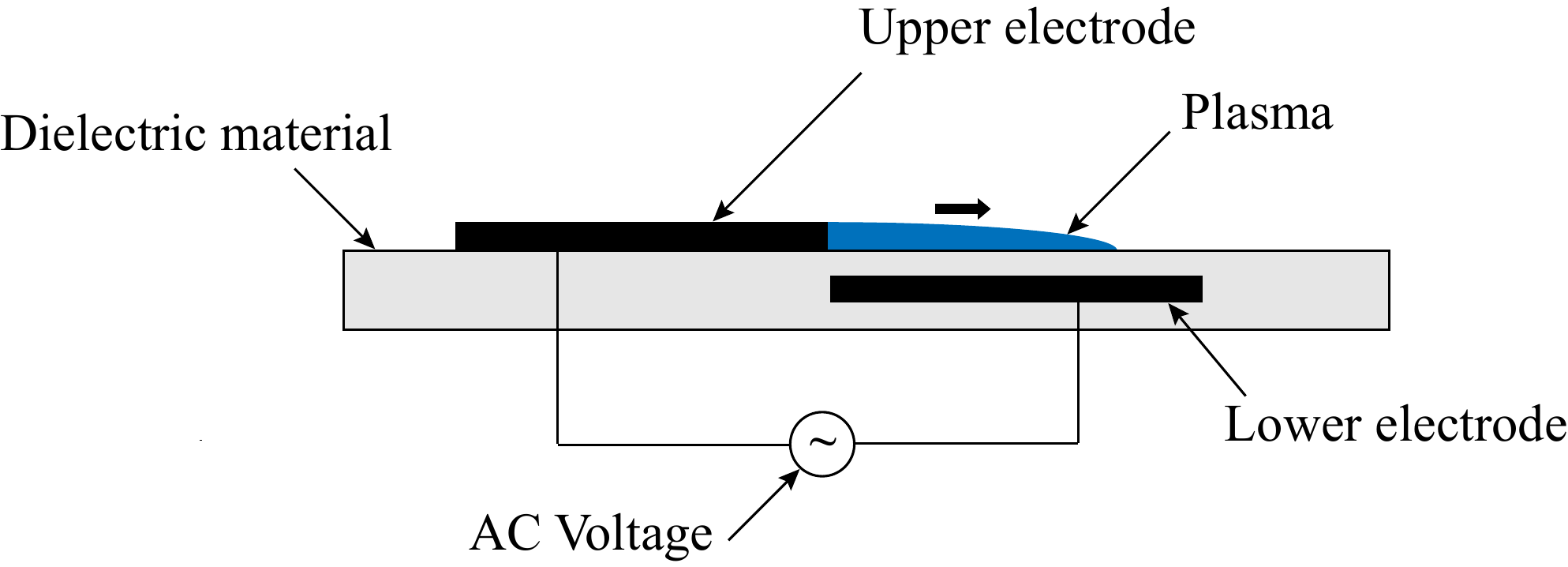}}
\caption[]{Configuration of a typical single DBD plasma actuator.}
\label{fig:1-PAs}
\end{figure}

 The configuration of a typical single plasma actuator consists of two electrodes separated by a dielectric layer, where the upper electrode is exposed to the fluid flow and the lower electrode is encapsulated, as illustrated in Figure~\ref{fig:1-PAs}. Plasma actuators work based on the principle of Dielectric Barrier Discharge (DBD), which forms a low-temperature plasma between the electrodes by applying high AC voltage. This electrical discharge from the electrodes ionizes the ambient flow, resulting in an induced flow towards the edge of the upper electrode in the direction of the lower electrode, generating momentum in the surrounding fluid and enabling control over the flow. For more details, refer to the reference papers.\cite{Benard2015, Orlov2008} \par

There are two main categories to model the effect of DBD plasma actuators numerically: first-principle-based and simplified phenomenological models. First-principle-based models established solutions of the complex transport equations for charged and neutral species, also a Poisson equation for the electric field. However, these highly precise models are computationally expensive compared to simplified ones and are not necessary for the purposes of this study. \cite{Boeuf2007, Nishida2016, Parent2016} Two commonly used simplified numerical models for plasma actuators are the Shyy model\cite{Shyy2002} and the Suzen model.\cite{Suzen2005, Suzen2007} The Shyy model is a simplified model that conducts computation for the glow discharge-induced fluid flow and heat transfer. The model uses basic features of the plasma physics along with an approximate modeling of the paraelectric effect on the flow. The model also accounts for the effect of the electric field on the flow, which is essential for understanding the behavior of plasma actuators. However, due to its simplicity, this model is considered less accurate regarding induced flow modeling. In contrast, the Suzen models, proposed by Suzen $et$ $al.$ in 2005 and 2007, are more detailed models that simulate the interaction between the plasma actuator and the flow field. The Suzen models are able to accurately predict the behavior of plasma actuators for a wide range of operating conditions and have been widely used. \par

For this study, the Suzen model proposed in 2007\cite{Suzen2007} is chosen to express the body force induced by the plasma actuator, because of its ability to eliminate the requirement to represent plasma distribution over the embedded electrode by Gaussian distribution as suggested by the model proposed in 2005\cite{Suzen2005}. The numerical model is derived assuming that magnetic forces are negligible, consequently, the electrohydrodynamic force can be defined as follows:

\begin{equation} \label{eqn:eq1}
{\bf f}=\rho_c{\bf E}
,
\end{equation}

\noindent where ${\bf f}$, $\rho_c$, and ${\bf E}$ are the body force, the charge density, and the electric field, respectively. \par

Maxwell's equation is expressed in terms of the gradient of a scalar electric potential $\Phi$, by assuming that time variation of the magnetic field is negligible such that

\begin{equation} \label{eqn:eq2}
{\bf E}= - \nabla \Phi
.
\end{equation}

\begin{figure}
\centering 
\scalebox{.6}{\includegraphics[angle=0, trim=0 0 0 0, width=0.8\textwidth]{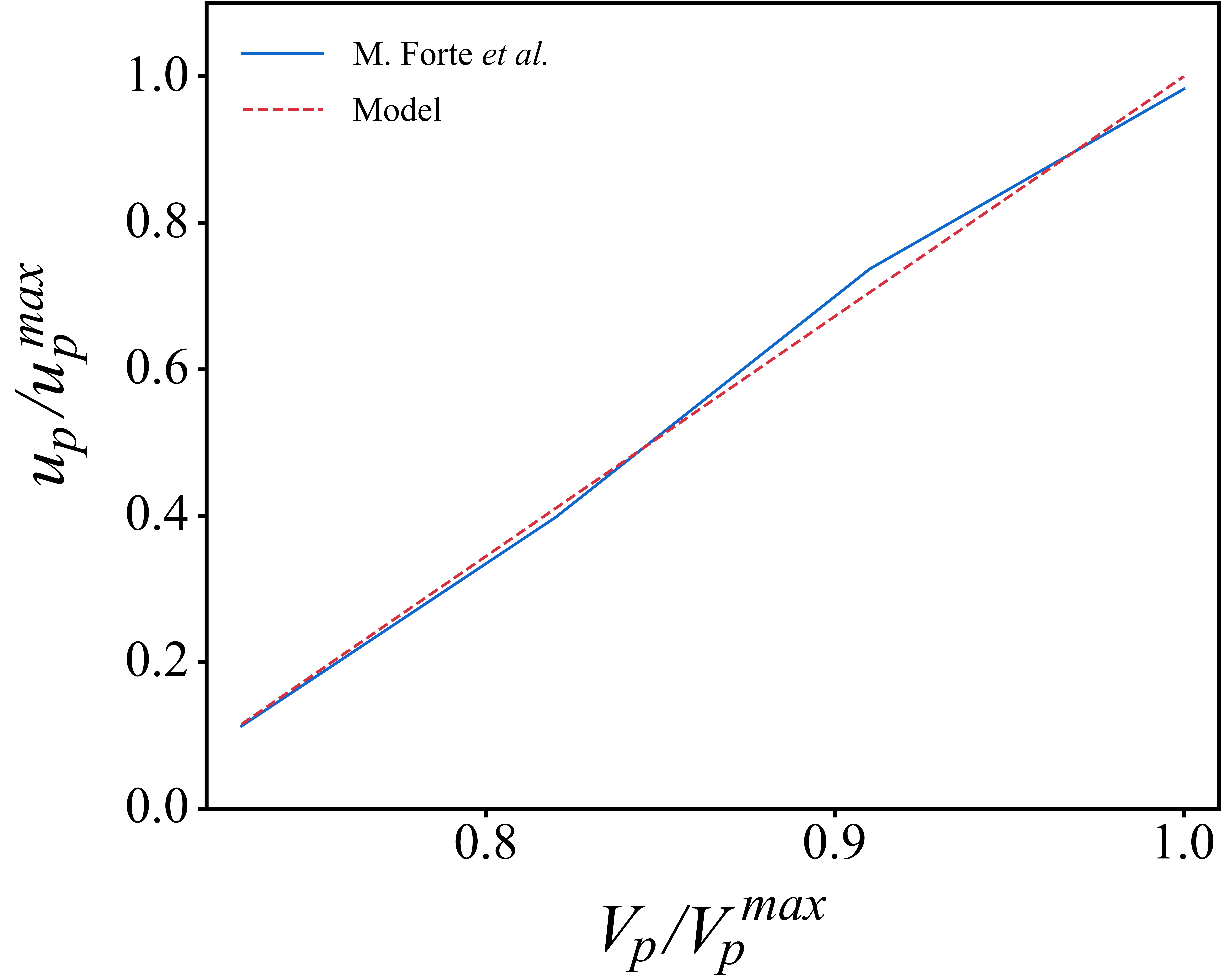}}
\caption[]{Induced velocity magnitude computed at the applied range of AC Voltage.}
\label{fig:2-CalibrationPAs}
\end{figure}

By using the Gauss's law $(\nabla\cdot(\varepsilon{\bf E)}=\rho_c)$, it can be

\begin{equation} \label{eqn:eq3}
\nabla\cdot\left(\varepsilon\nabla\Phi\right)=-\rho_c
,
\end{equation}
\begin{equation} \label{eqn:eq4}
\varepsilon=\varepsilon_r\varepsilon_0
.
\end{equation}

\noindent where $\varepsilon, \varepsilon_r$, and $\varepsilon_0$ are the permittivity, relative permittivity, and free-space permittivity, respectively. The electric potential can be split into the electric potential due to the external electric field $\phi$, and the electric potential due to the net charge density, $\varphi$. Equation (3) can be replaced by two separate equations,
	
\begin{equation} \label{eqn:eq5}
\nabla\cdot\left(\varepsilon_r\nabla\phi\right)=0
,
\end{equation}

\begin{equation} \label{eqn:eq6}
\nabla\cdot\left(\varepsilon_r\nabla\varphi\right)=-\frac{\rho_c} {\varepsilon_0}
. 
\end{equation}

\begin{figure}
\centering 
\scalebox{1.2}{\includegraphics[angle=0, trim=0 0 0 0, width=0.8\textwidth]{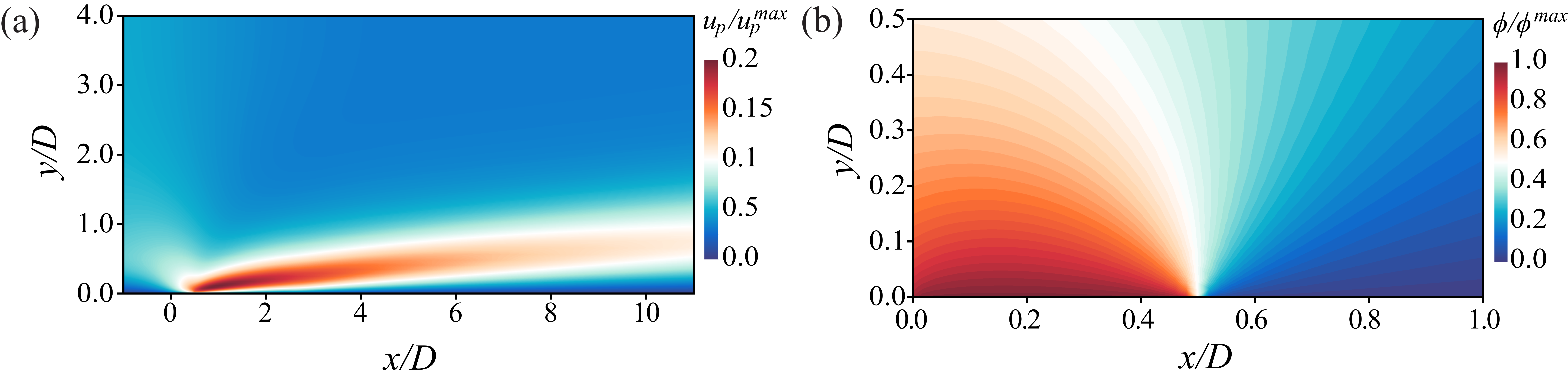}}
\caption[]{Instantaneous contours of a typical single DBD plasma actuator at $V_{p}{/V}_{p}^{max} =0.75$: (a) streamwise velocity field and (b) electric potential field.}
\label{fig:3-contoursPAs}
\end{figure}

A relationship between $\varphi$ and $\rho_c$ can be established such that:

\begin{equation} \label{eqn:eq7}
\varphi=-\frac{\rho_c\lambda_d^2}{\varepsilon_0}
, 
\end{equation}

\noindent where $\lambda_d$ is the Debye length, which is the characteristic length for electrostatic shielding in plasma.
Equation (6) can be re-written as,

\begin{equation} \label{eqn:eq8}
\nabla\cdot\left(\varepsilon_r\nabla\rho_c\right)=-\frac{\rho_c}{\lambda_d^2}
. 
\end{equation}
 
By combining equations (5) and (8), the body force can be obtained as,

\begin{equation} \label{eqn:eq9}
{\bf f}=\rho_c{\bf E}=\rho_c(-\nabla\phi)
. 
\end{equation}
 
As equations (5) and (8) are time-independent, the normalized equations can be solved at the beginning of the simulation. To determine the current body force, the maximum values of the normalized electric potential and charge density are multiplied by their respective normalized values. This allows for the determination of the appropriate body force in the simulation as: 

\begin{equation} \label{eqn:eq10}
{\bf f}={(-\nabla\phi^\ast\left(x\right))\ \phi}^{max}{\rho_c}^\ast(x){\rho_c}^{max}\
, 
\end{equation}

\noindent where $\rho_c^\ast$ and $\phi^\ast$ are the charge density and electric potential normalized by their maximum values, i.e., $\rho_c^{max}$ and $\phi^{max}$. The values of  ${\rho_c}^{max}$ and $\lambda_{d}$ are obtained empirically referring to the experimental results of Forte $et$ $al.$\cite{Forte2007} A comparison of the obtained results is illustrated in Figure~\ref{fig:2-CalibrationPAs}. We use a local linear approximation of the maximum induced velocity $u_p$ generated by a typical single DBD plasma actuator provided in reference paper with a fixed frequency of applied AC voltage ($f=1kHz$) at a range $V_{p}\in\ [0.7;\ 1.0]$$V_{p}^{max}$, where $V_{p}^{max}=11kV$ is the maximum applied voltage with the corresponded maximum induced velocity, $u_{p}^{max}=2 m/s$. The instantaneous velocity and electric potential contours generated by a typical single DBD plasma actuator at $V_{p}{/V}_{p}^{max} =0.75$ are illustrated in Figure~\ref{fig:3-contoursPAs}. \par

\subsection{Details of simulation environment}

A two-dimensional direct numerical simulation (DNS) is conducted to analyze the incompressible flow over the square cylinder. The continuity and momentum equations are represented as follows:
\begin{equation} \label{eqn:eq11}
\mathrm{\nabla}\cdot {\bf u} = 0
, 
\end{equation}
\begin{equation} \label{eqn:eq12}
\frac{\partial {\bf u}} {\partial t} + \nabla\cdot ({\bf u} {\bf u})=-\nabla p + \nabla \cdot (\nu \nabla {\bf u}) + {\bf f} , 
\end{equation}

\noindent where ${\bf u}$ is the velocity vector, $t$ is the time step, $p$ is a pressure per density, $\nu$ is the kinematic viscosity, and  ${\bf f}$  is the source term body force used to capture the ionization effects of the plasma actuators. More detailed body force derivation equations are explained in subsection \ref{subsec:II2}. The Reynolds number based on the free-stream velocity $U_\infty$, and the cylinder diameter $D$ is ${Re}_D=\frac{U_\infty D}{\nu}$, where $\nu=1.5\times{10}^{-5}m/s^2$; $Re_D$ =100 and 180 are used in this study. The drag coefficient $C_D$, lift coefficient $C_L$, pressure coefficient $C_p$, Strouhal number $St$, and dimensionless wall distance $y^+$ are defined as $C_D=\ \frac{F_D}{\frac{1}{2}\rho U_\infty^2D^2}$, $C_L=\ \frac{F_L}{\frac{1}{2}\rho U_\infty^2D^2}$, $C_p=\ \frac{{p-p}_\infty}{\frac{1}{2}\rho U_\infty^2}$, $St=\ \frac{fD}{U_\infty}$, and $y^+=\frac{yu_\tau}{\nu}$  respectively, where $F_D$ is the drag force, $F_L$ is the lift force, $p_\infty$ is the free-stream pressure, $f$ is the vortex shedding frequency, $u_\tau$ is the friction velocity, $y$ is the absolute distance from the wall, and $\rho$ is the density.\par 

\begin{table}
  \begin{center}
  \caption{Detailed information of computational domain}
\scalebox{1.0}{
\begin{tabular}{ccccccccc} \hline\hline
&Mesh~ ~&~~Number of cells~ ~&~~Number of points~ ~&~~Wall-normal expansion ratio~ ~&~~ Maximum $y^+$& \\ \hline
&Coarse~ ~&~~$27492$~~&~~$56160$~~&~~$1.15$~~&~~$0.40$& \\
&Medium~ ~&~~$35200$~~&~~$71652$~~&~~$1.12$~~&~~$0.37$& \\
&Fine~ ~&~~$48656$~~&~~$98692$~~&~~$1.10$~~&~~$0.37$& \\  \hline\hline
\end{tabular}}
  \label{tab:table1}
  \end{center}
\end{table}

\begin{figure}
\centering 
\scalebox{.7}{\includegraphics[angle=0, trim=0 0 0 0, width=0.8\textwidth]{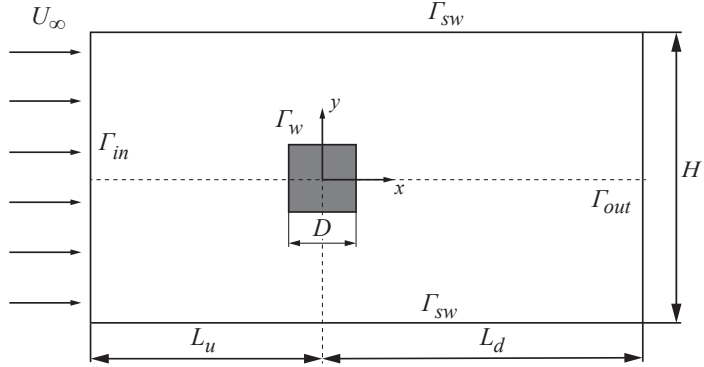}}
\caption[]{Schematic of the computational domain.}
\label{fig:4-Domain}
\end{figure}

The schematic of the computational domains is illustrated in Figure~\ref{fig:4-Domain}. The dimensions of the model in the streamwise and lateral directions are set to be  $L_x=25D$ and $L_y=20D$, respectively. The distances of the upstream and downstream boundaries from the center of the cylinder are $L_u=9D$ and $L_d=16D$. The computational simulations are conducted using a structured mesh, which is refined in the vicinity of the cylinder, resulting in a total of 35,200 cells. The time-averaged coefficients are obtained for $20s$ of the simulation time after it reaches convergence. The parameters of the mesh are validated, and its details are provided in Table~\ref{tab:table1}. Furthermore, a comparison of the simulation results with those obtained from reference papers is presented in Table~\ref{tab:table2}. \par

For boundary conditions, the slip $\Gamma_{sw}$ boundary condition is assigned to the top and bottom sides of the computational domain, while the no-slip $\Gamma_{w}$ condition is applied to the cylinder walls. Uniform inlet velocity $\Gamma_{in}$ and pressure outlet $\Gamma_{out}$ boundary conditions are employed at the inlet and outlet of the domain, respectively. The numerical model of the plasma actuators described in subsection \ref{subsec:II2} is integrated into the finite volume method (FVM) based open-source CFD code, OpenFOAM-6.0, and coupled with a solver based on the PIMPLE algorithm. The numerical simulation is conducted by incorporating two iterative loops for the pressure-velocity coupling, and the time step is $1\times{10}^{-4} s$.\par

\begin{table}
  \begin{center}
    \caption{Validation of various meshes.}
    \setlength{\tabcolsep}{6pt}
\scalebox{1.0}{
\begin{tabular}{ccccccccc} \hline\hline
&Case~ ~&~~$C_{L,rms}$~ ~&~~$\overline{C_D}$~ ~&~~$St$& \\ \hline
&Coarse~ ~&~~$0.174$~~&~~$1.495$~~&~~$0.144$& \\
&Medium~ ~&~~$0.173$~~&~~$1.495$~~&~~$0.145$& \\
&Fine~ ~&~~$0.173$~~&~~$1.495$~~&~~$0.145$& \\
&Sohankar $et$ $al.$\cite{Sohankar1998}~ ~&~~$0.156$~~&~~$1.477$~~&~~$0.146$& \\  
&Sahu $et$ $al.$\cite{Sahu2009}~ ~&~~$0.188$~~&~~$1.488$~~&~~$0.149$& \\
&Sen $et$ $al.$\cite{Sen2011}~ ~&~~$0.193$~~&~~$1.530$~~&~~$0.145$& \\ \hline\hline
\end{tabular}}
  \label{tab:table2}
  \end{center}
\end{table}

\subsection{Deep Reinforcement Learning Framework}

Active flow control encounters significant challenges due to the complex and high-dimensional nature of fluid flows. Non-linear governing equations make it difficult to develop effective control strategies, while the dynamic and complex behaviors of the flow further complicate the process. To overcome these challenges, we use a DRL model based on an artificial neural network (ANN) architecture. Deep Reinforcement Learning (DRL) is is a type of machine learning that considers an agent's interactions with its environment through three channels: state ($s_t$), action ($a_t$), and reward ($r_t$). In our study, the system's states are represented by an array of pressure probes distributed throughout the domain. The agent's actions correspond to the regulation of the AC voltage amplitude applied to the plasma actuators, and the reward function considers the combination of the drag and lift coefficients obtained after each action period ($\Delta t$). The agent's goal is to learn a policy that maximizes the cumulative reward over time, aligning with desired flow control outcomes. Our ANN consists of 512 fully connected neurons, enabling the learning of approximately 300,000 parameters. The choice of the number of neurons is based on a study conducted by Rabault $et$ $al$.\cite{Rabault2019_1} And the proximal policy optimization (PPO) algorithm\cite{Schulman2017} is employed for the agent to learn a flow control strategy.\cite{Rabault2019_1, Li2021, Wang2022}\par

To mitigate computational costs during the training process, the duration of each applied action $\Delta t$ is set to a constant value corresponding to 4.5\% of the vortex shedding cycle, which is lower than the action duration used by Rabault $et$ $al$.\cite{Rabault2019_1} The parameters used in this study are found to be effective for the current flow control problem, resulting in 10 taken actions for each training episode, which is equivalent to approximately half of the vortex-shedding cycle for the flow at ${Re}_D=100$. To update the networks, the adam optimizer is employed, and the learning rate is set to a fixed value of $1\times{10}^{-3} s$. The ANN is trained to minimize $C_D$ and $C_L$, thus maximizing reward function $r_t$, defined as follows:

\begin{equation} \label{eqn:eq13}
r_t=-({|C}_D|+\alpha\left|\left\langle C_L\right\rangle_T\right|)\
, 
\end{equation}

\noindent where $\langle\cdot\rangle_T$ represents the moving average of the lift coefficient back in time over a duration that is equal to one vortex shedding cycle, and $\alpha$ is the penalization coefficient required to balance the sensitivity of the training model from the drag and lift coefficients, defined as 0.2. The inclusion of the absolute value of the drag coefficient in the reward function aims to prevent the DRL model from converging to negative values of the drag. In the case of square cylinders, a negative drag coefficient would imply that the thrust effect from the actuation is higher than the drag generated by the flow around the cylinder. In such situations, the cylinder may experience a net thrust force in the direction of the actuation, leading to a negative effective drag coefficient. \par

\begin{figure}
\centering 
\scalebox{0.8}{\includegraphics[angle=0, trim=0 0 0 0, width=0.9\textwidth]{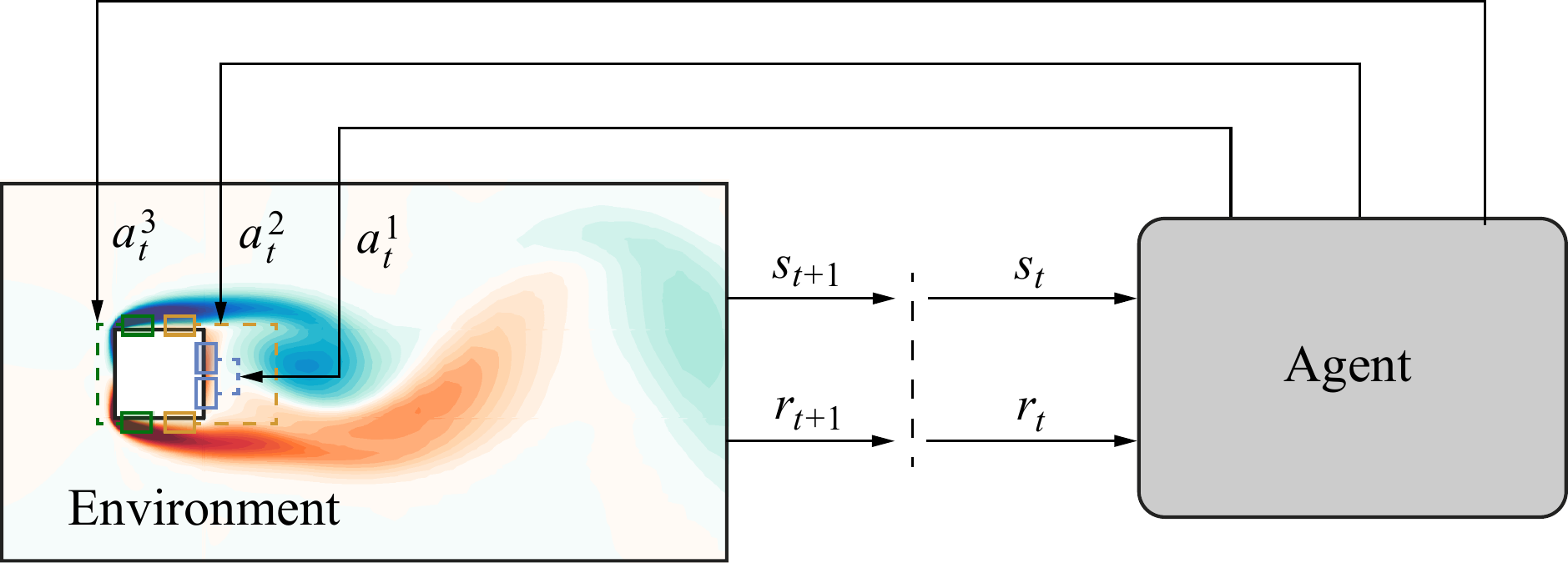}}
\caption[]{Schematic of reinforcement learning loop.}
\label{fig:5-DRL-loop}
\end{figure}

\begin{figure}
\centering 
\scalebox{0.6}{\includegraphics[angle=0, trim=0 0 0 0, width=0.9\textwidth]{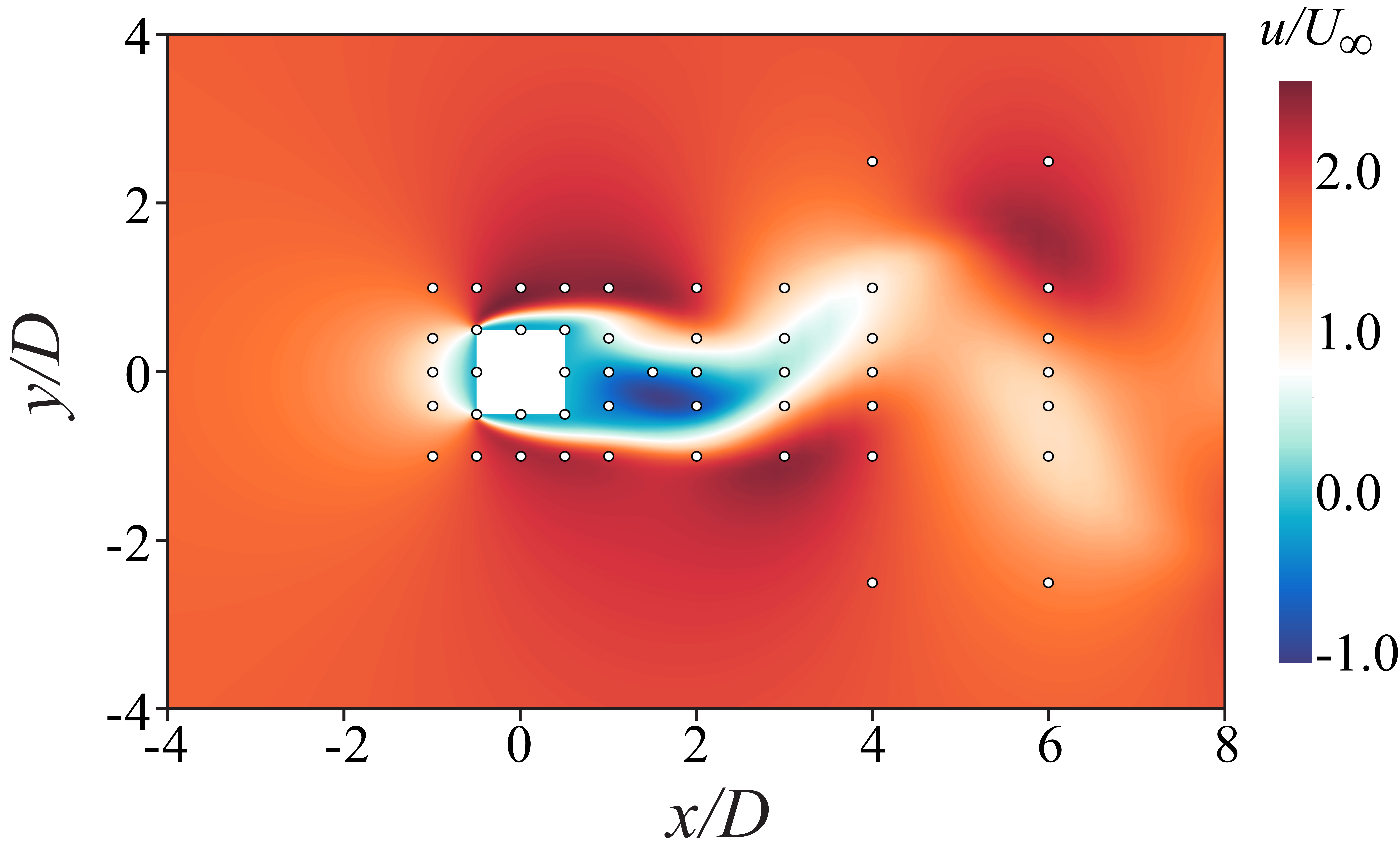}}
\caption[]{Location of the sampled probes.}
\label{fig:6-probes}
\end{figure}

The strategic spatial distribution of the probes is crucial in capturing flow dynamics that significantly affect the control objectives. As mentioned before, the state observation is represented by the distribution of the pressure probes and involves the identification of the regions of high flow gradients, formation of vortices, and flow separation. Therefore, after thorough testing and experimentation, we have determined that a total of 49 probes, carefully allocated in the domain, can provide informative data for the DRL agent. It is important to note that too few probes may result in sparse data, leading to inaccurate flow field estimations and limited learning for the DRL agent. Conversely, an excessive number of probes may lead to redundant data and increase computational costs. The distribution of pressure probes for the current study is illustrated in Figure~\ref{fig:6-probes}. Furthermore, in subsection \ref{subsec: B2}, a detailed analysis is conducted to investigate the specific contribution of the pressure probes to the model.\par

\begin{figure}
\centering 
\includegraphics[angle=0, trim=0 0 0 0, width=0.9\textwidth]{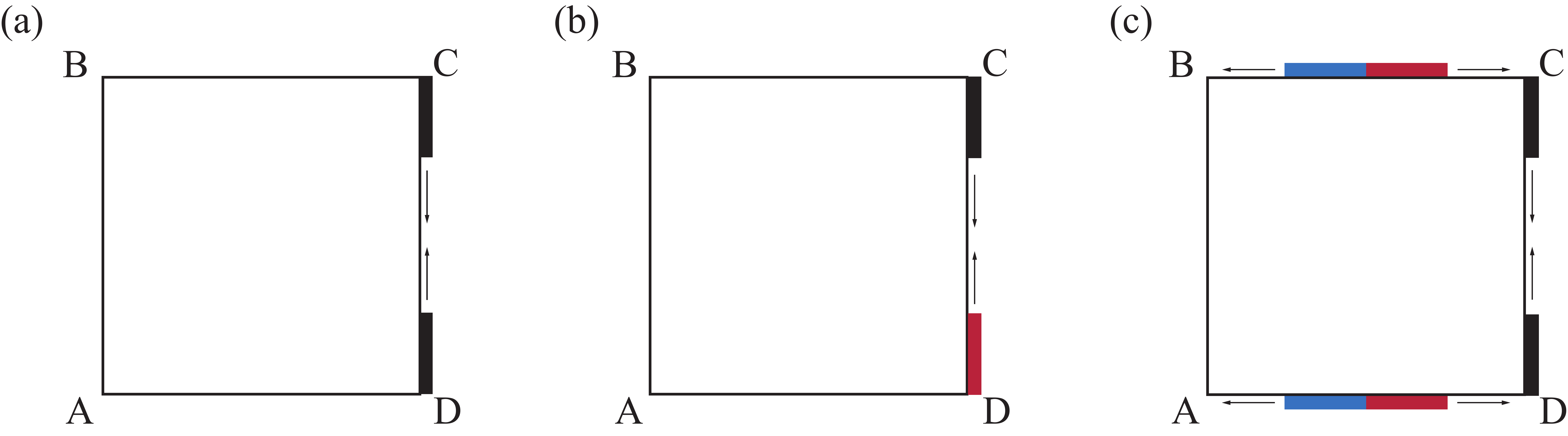}
\caption[]{Schematic depicting the placement of the plasma actuators. (a) Case 1; (b) Case 2; (c) Case 3.}
\label{fig:7-PAs-placement}
\end{figure}

In this study, the training process is conducted across three distinct configurations of the plasma actuators at Reynolds numbers (${Re}_D$) of $100$ and $180$ separately, as shown in Figure~\ref{fig:7-PAs-placement}, where the shown rectangles on the cylinder surface represent the schematic allocation of the upper electrodes of the plasma actuators. The first configuration (Case 1) involved training with a single pair of plasma actuators positioned at the rear surface of the square cylinder,  sourced from a single AC signal ($a_t$). The second configuration (Case 2) is similar to the first, however, plasma actuators are sourced from two different AC signals ($a_t^{1}$ and $a_t^{2}$). Lastly, the third configuration (Case 3) encompassed training with three pairs of plasma actuators, positioned at the rear surface of the cylinder ($a_t^{1}$), trailing top-bottom surfaces ($a_t^{2}$), and leading top-bottom surfaces ($a_t^{3}$). Additionally, a continuous control method is employed to update the control for each action based on the previous amplitude of the AC voltage in order to ensure smooth control and prevent simulation instability. Specifically, the new control ($\hat{a}_{t}$) is determined using the formula $\hat{a}_{t}=\hat{a}_{t-\Delta t}+\beta$($a_t-\hat{a}_{t-\Delta t}$), where $\hat{a}_{t-\Delta t}$ represents the amplitude of AC voltage used in the previous action period, $a_t$ is an action determined by the PPO-agent, and $\beta$ is a constant defined as 0.1. The computed range of action space is defined as $a_t\in [0.7,1.0]V_{p}^{max}$. \par

\section{Results and discussion}

\begin{figure}
\centering 
\includegraphics[angle=0, trim=0 0 0 0, width=0.9\textwidth]{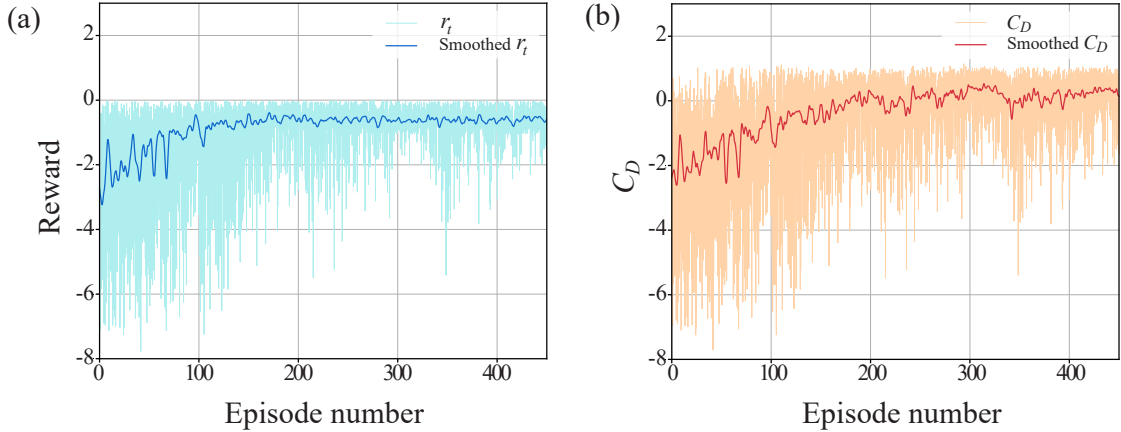}
\caption[]{Illustration of the training process: progress of instantaneous (a) reward and (b) drag coefficient.}
\label{fig:8-training}
\end{figure}

The resilience of the learning process is demonstrated in Figure~\ref{fig:8-training}. The training process spans 450 episodes, where the reward function achieves convergence within 200 episodes. Notably, it is relevant to highlight that the control strategy in this study is slightly different from that observed for circular cylinders. Specifically, the time evolution of the drag coefficient during the training process exhibits an increment from negative to zero, in contrast to the opposite behavior observed for the circular cylinder\cite{Rabault2019_1}. This observation is closely related to the square cylinder's control strategy using plasma actuators, whereby the cylinder may experience a negative drag value when the actuation thrust effect is greater than the drag generated by the flow around it. \par

First, the effectiveness of the proposed model is investigated by conducting tests at ${Re}_D=100$ and further applied for a case at ${Re}_D=180$. All quantities are non-dimensionalized by the time duration of one action $\Delta t$, the cylinder diameter $D$, the free-stream velocity $U_\infty$, maximum induced velocity $u_p^{max}$, and maximum amplitude of applied AC voltage  $V_p^{max}$. The uncontrolled case is denoted as NC and the first 20 shedding cycles of the NC case are ignored to neglect the transitional period. \par

Table~\ref{tab:table3} presents a comparison of the control performance between the DRL-based approach and other studies that utilized plasma actuators to control the flow past a square cylinder. The results indicate that the DRL-based approach outperformed the reference study results in terms of finding the optimal level of applied actuation and eventually reducing both drag and lift forces. Moreover, to evaluate the effect of the drag reduction obtained from the suppression of the vortex shedding we use a corrected drag coefficient $\Delta C_{D-cor.}$, defined as:

\begin{equation} \label{eqn:eq14}
\Delta C_{D-cor.} = \Delta C_D - C_T
, 
\end{equation}

\begin{table} 
\begin{center}
~\caption{Comparison of the control performance.}
\setlength{\tabcolsep}{8pt}
\scalebox{1.0}{
\begin{tabular}{ c c c c c c c c c c c c}
\hline\hline
& $Re_D$ & Case & & $C_{L,rms}$ &~~$\overline{C_D}$~~& $\Delta C_D/\overline{C_D}$ & $C_T$ & $\Delta C_{D-cor.}/\Delta C_D$ &	\\ \hline
& & NC & & 0.173 & 1.495 & & & &  		\\ 
& & 1 & & 0.000 & 0.211 & 0.86 & 0.698 & 0.46 &		\\ 
& 100 & 2 & & 0.000 & 0.107 & 0.93 & 0.765 & 0.457		\\ 
& & 3 & & 0.000 & 0.009 & 0.97 & 0.624 & 0.58 &		\\ 
& & Zhu et al.\cite{Zhu2021} & & 0.000 & 0.400 & & & &  		\\ 
& & Anzai et al.\cite{Anzai2017} & & - & 0.700 & & & &  		\\ \hline
& & NC & & 0.382 & 1.483 & & & & 		\\ 
& 180 & 1 & & 0.086 & 0.273 & 0.82 & 1.004 & 0.17 & 		\\ 
& & 3 & & 0.000 & 0.002 & 0.99 & 1.146 & 0.23 &		\\ 
& 200 & Z. Zhu, et al.\cite{Zhu2021} & & 0.170 & 1.800 & & & &  		\\ \hline \hline
\end{tabular}} \label{tab:table3}
\end{center}
\end{table}

\noindent where $\Delta C_D$ is the total reduction of the mean drag coefficient and $C_T$ is the coefficient of the total thrust $T$ of the plasma actuators. The trust effect of the plasma actuators computed in the quiescent air, defined as $C_T=\ \frac{T}{\frac{1}{2}\rho U_\infty^2D^2}$. The results show that the contribution of the vortex shedding elimination to the drag reduction decreases as the Reynolds number increases. Conversely, due to the relatively low flow complicity at ${Re}_D = 100$, the use of plasma actuation with thrust ratio ($C_T/\Delta C_D\approx0.5$) is found to be sufficient to suppress the shedding vortices. This is evidenced by the measured suppression of lift fluctuations in the system. In contrast, a higher thrust ratio ($C_T/\Delta C_D\approx0.8$) is required to effectively suppress the shedding and reduce lift fluctuations for the case at ${Re}_D = 180$.\par 

\subsection{Effectiveness of the DRL model at $\bf Re_D = 100$}

\begin{figure}
\centering 
\scalebox{1.1}{\includegraphics[angle=0, trim=0 0 0 0, width=0.9\textwidth]{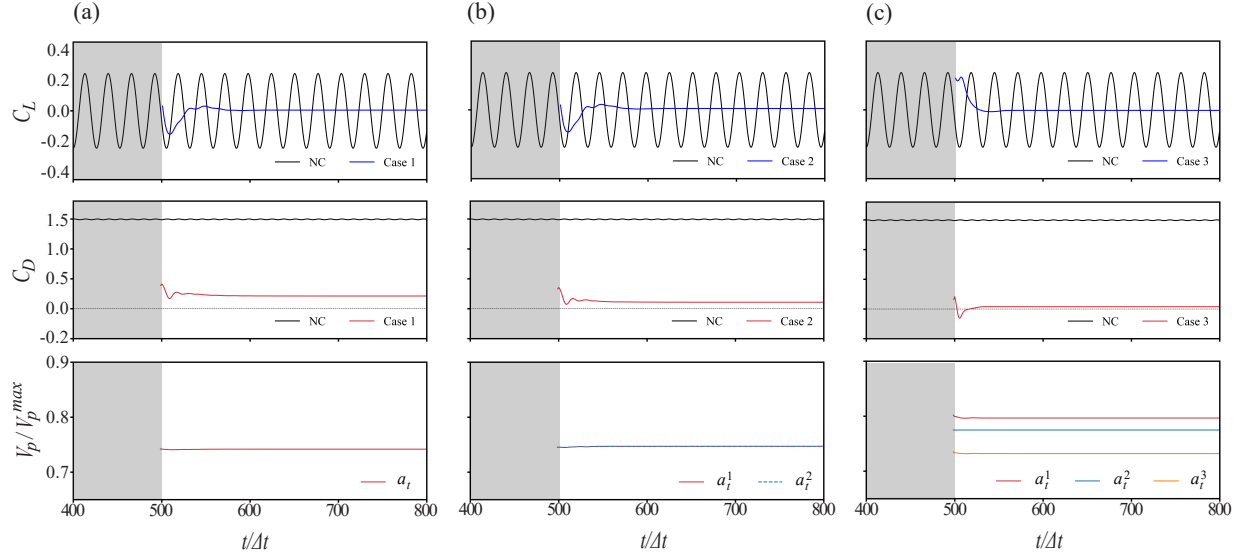}}
\caption[]{Evolution of the instantaneous drag coefficient, lift coefficient, and applied actions at ${Re}_D=100$: (a) Case 1; (b) Case 2; (3) Case 3.}
\label{fig:9-control100}
\end{figure}

The results of the applied active flow control strategy for Case 1, Case 2, and Case 3 are presented in Figure~\ref{fig:9-control100}, with separate columns for each case. The gray background in each column represents the flow initialization period, where the flow reaches quasi-steady conditions before the actuation is applied. The outcome shows that the application of the active flow control strategy leads to significant reductions in the drag coefficients for all cases, with reductions of 86\%, 93\%, and 97\% observed for Case 1, Case 2, and Case 3, respectively. Additionally, the applied control strategy is able to completely suppress lift oscillations across all cases. \par

Note that the amplitude of the applied action chosen by the first model can completely suppress lift fluctuations after 75 non-dimensional time steps of control, which is also observed for the second model. The temporal evolution of lift and drag coefficients shows similar behavior in both cases, indicating comparability of their performance. Although the agent of Case 2 trained with two actions, it converged to a solution where the same level of actuation is applied for both plasma actuators. On the other hand, the control strategy selected by the third model is able to achieve complete suppression of lift fluctuations after 50 non-dimensional time steps of control, involving three different actions. Action 1 is converged to the highest amplitude and involves a pair of plasma actuators positioned on the rear surface of the cylinder. Action 2 has a middle amplitude and is applied to the pair of plasma actuators located near the trailing edge of the cylinder's top and bottom surfaces. Action 3 is corresponded to the lowest amplitude and involves the pair of plasma actuators placed close to the leading edge of the cylinder's top and bottom surfaces. \par

\begin{figure}
\centering 
\scalebox{1.1}{\includegraphics[angle=0, trim=0 0 0 0, width=0.9\textwidth]{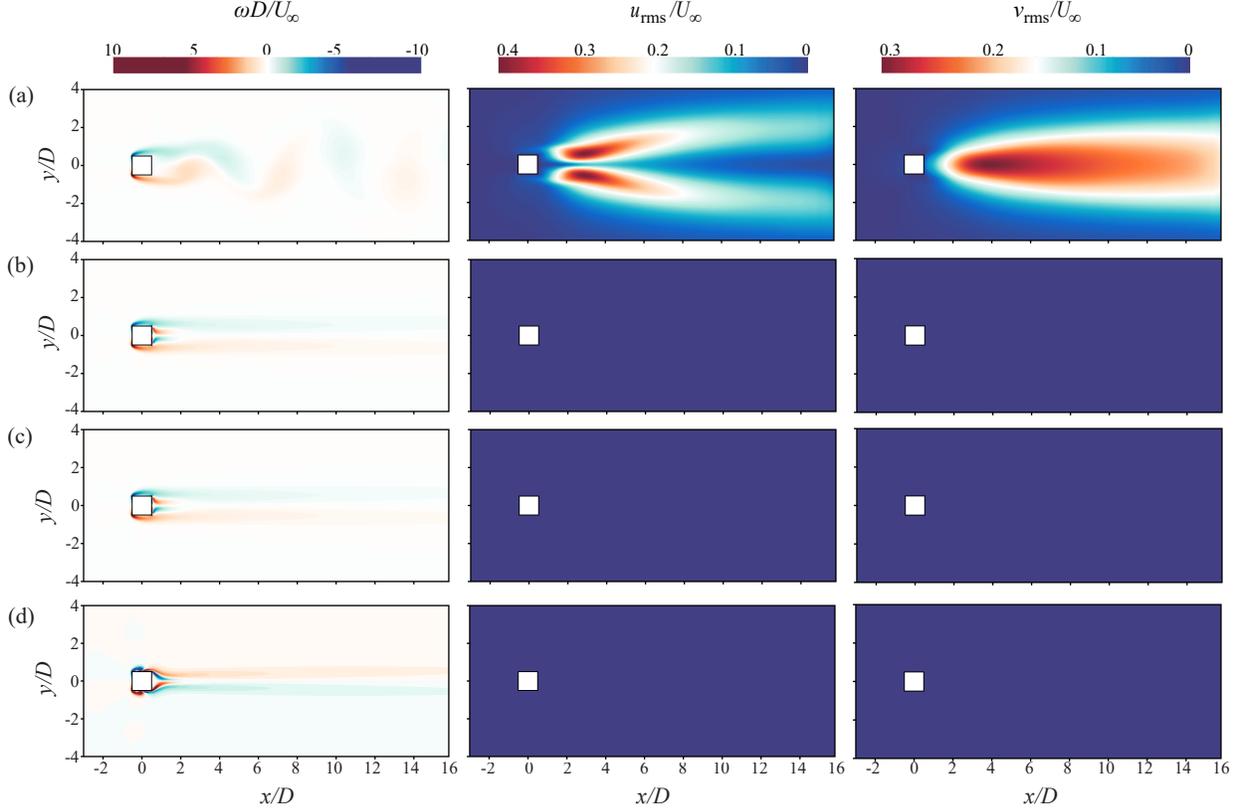}}
\caption[]{Instantaneous vorticity, rms streamwise velocity fluctuations, and rms transverse velocity fluctuations distributions at ${Re}_D=100$: (a) NC; (b) Case 1; (c) Case 2; (d) Case 3.}
\label{fig:10-contour100}
\end{figure}

The results presented in Figure~\ref{fig:10-contour100} display contours of the instantaneous vorticity ($\omega$), root-mean-square (rms) streamwise velocity fluctuations ($u_{rms}$), and rms transverse velocity fluctuations ($v_{rms}$), before and after activating the control. The flow field for the uncontrolled case is represented by a repeating pattern of a von Karman vortex shedding. When the control is applied the shedding of vortices is observed to be completely suppressed for all cases. This behavior is similar to the one observed by Anzai $et$ $al.$,\cite{Anzai2017} for a case with plasma actuators allocated at the rear surface of the square cylinder at ${Re}_D=100$. In cases 1 and 2, the effects of the two actuators are results in a vortices near the rear surface of the cylinder. In Case 3, a pair of forced vortices are observed to be allocated at the top and bottom surfaces. Remarkably, both the streamwise velocity fluctuations and transverse velocity fluctuations exhibit a significant reduction and essentially vanish for all cases, indicating a highly effective control strategy. Further, due to the similarity in behavior and comparable performance of the control strategy for cases 1 and 2, further analysis is conducted for only cases 1 and 3 for the flow at ${Re}_D=180$. \par

\subsection{Effectiveness of the DRL model at $\bf Re_D = 180$}\label{subsec: B2}

\begin{figure}
\centering 
\scalebox{0.9}{\includegraphics[angle=0, trim=0 0 0 0, width=0.9\textwidth]{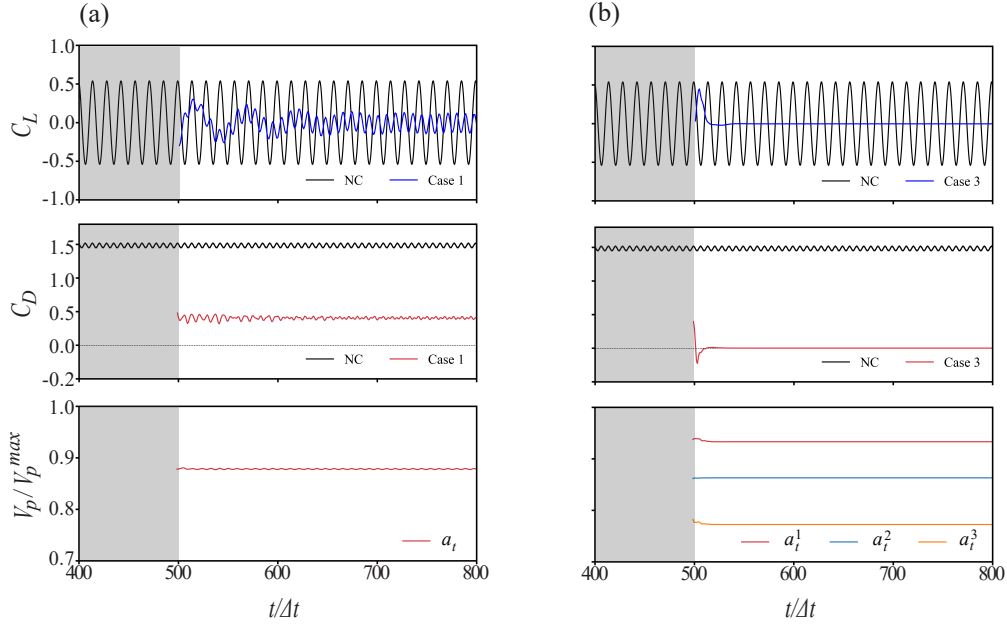}}
\caption[]{Evolution of the instantaneous drag coefficient, lift coefficient, and applied actions at ${Re}_D=180$: (a) Case 1; (b) Case 3.}
\label{fig:11-controll180}
\end{figure}

The performance of the AFC applied for a case at ${Re}_D=180$ is illustrated in Figure~\ref{fig:11-controll180}. The findings demonstrate that as the Reynolds number increases, suppressing two-dimensional vortex shedding with plasma actuators located only on the rear surface of the cylinder (Case 1) becomes challenging. Despite this, for Case 1 the mean drag coefficient is observed to decrease by 82\%, indicating a significant reduction. However, there is a slight increment in the amplitude of its fluctuations, mainly caused by the oscillatory behavior of the applied action. In terms of lift fluctuations, the control strategy can reduce them by 80\%, although complete suppression is not achieved. On the other hand, for Case 3, the control strategy chosen by the agent is able to achieve a mean drag coefficient reduction of 99\% and complete suppression of the lift fluctuations. \par

Based on the analysis results, the control strategy selected by the agent for Case 3, at a Reynolds number of 180, closely resembles that of Case 3 at a Reynolds number of 100. In this context, action 1 corresponds to the highest AC voltage amplitude applied, action 2 refers to the middle range, and action 3 pertains to the lowest amplitude. \par

\begin{figure}
\centering 
\scalebox{1.1}{\includegraphics[angle=0, trim=0 0 0 0, width=0.9\textwidth]{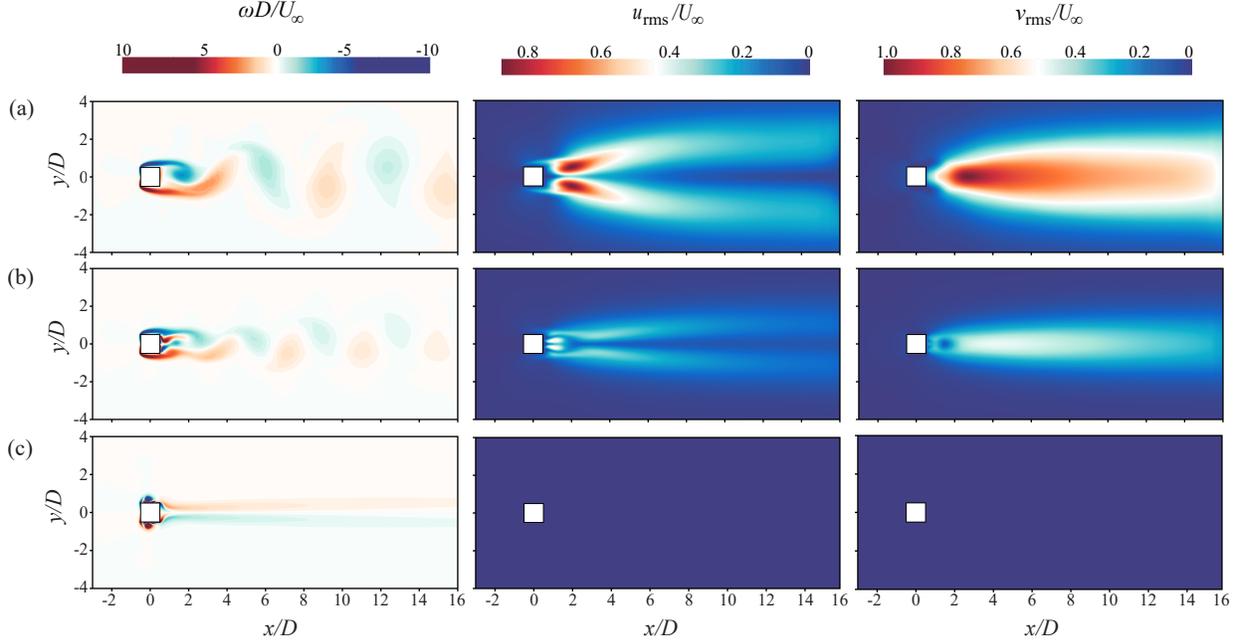}}
\caption[]{Instantaneous vorticity, rms streamwise velocity fluctuations, and rms transverse velocity fluctuations, distributions at ${Re}_D=180$ : (a) NC; (b) Case 1; (c) Case 3.}
\label{fig:12-contour180}
\end{figure}

The instantaneous contours demonstrated in Figure~\ref{fig:12-contour180} are in good agreement with results shown in Figure~\ref{fig:11-controll180}. The velocity induced by plasma actuators is able to modify the flow behavior, resulting in a comparable reduction of the vortex shedding for Case 1 and its complete elimination for Case 3. This observation is also visible from the rms of the streamwise and transverse velocity fluctuations.\par

\begin{figure}
\centering 
\includegraphics[angle=0, trim=0 0 0 0, width=0.9\textwidth]{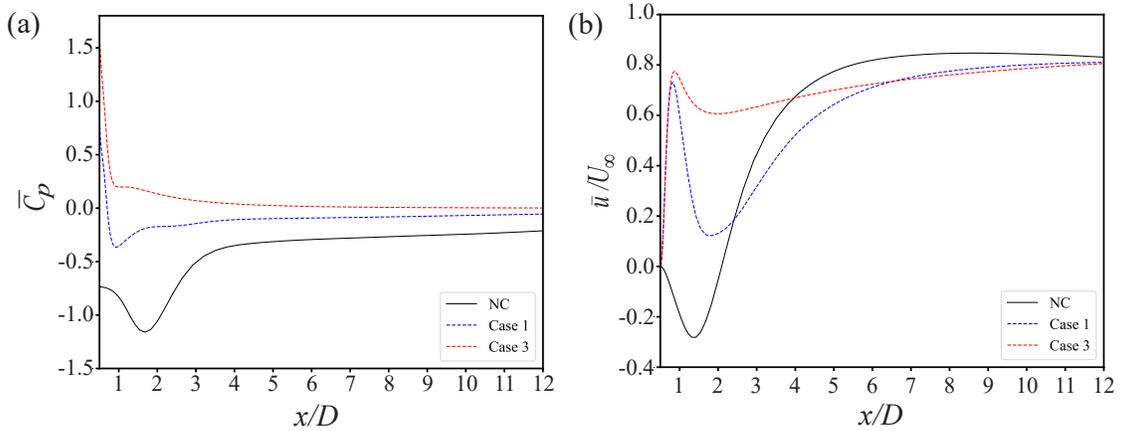}
\caption[]{Statistics on the wake centerline at ${Re}_D = 180$: (a) mean pressure coefficient; (b) mean streamwise velocity.}
\label{fig:13-centerline}
\end{figure}

The statistical analysis of the wake centerline, as illustrated in Figure~\ref{fig:13-centerline}, is revealed to be very similar to those observed in the case of manual controlling using plasma actuators.\cite{Anzai2017} Consequently, the pick of the centerline variation of the mean streamwise velocity at $x/D=1$ for controlled cases demonstrates the induced velocity by plasma actuators allocated at the rear surface. It is important to note that the pressure coefficient $\overline{C_p}$ at the rear surface in the NC case exhibits a negative value, which is also evident from the mean streamwise velocity profile $\bar{u}/U_\infty$ at the wake centerline, indicating flow separation and presence of a recirculation region in the wake. The centerline variation of the mean pressure coefficient, for controlled Case 1, shows that the central recirculation zone is moved upstream close to the rear surface of the cylinder and significantly suppressed its adverse pressure gradient. In addition, the most significant change has been found for Case 3, where the pressure coefficient is found to be a positive value, denoting that the recirculation region is totally eliminated. \par

\begin{figure}
\centering 
\includegraphics[angle=0, trim=0 0 0 0, width=0.9\textwidth]{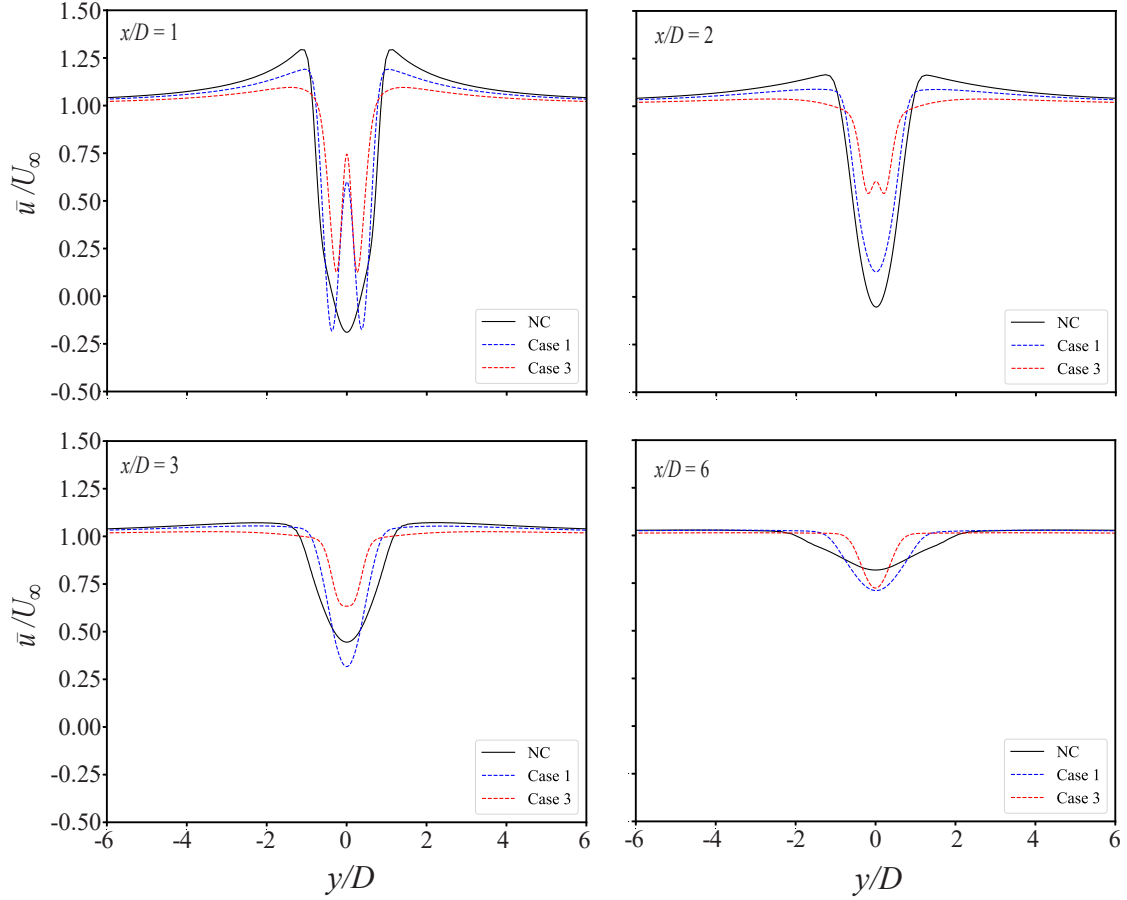}
\caption[]{Statistics statistics of the mean streamwise velocity for four locations in the wake at $x/D$ = 1, 2, 3, and 6 at ${Re}_D = 180$.}
\label{fig:14-span}
\end{figure}

This is also evident from statistics of the mean streamwise velocity for four locations in the wake at $x/D=1,2,3$, and $6$, as shown in Figure~\ref{fig:14-span}, where the value of the streamwise velocity component remains positive and does not reach negative values for Case 3. Additionally, interesting behavior can be observed in terms of flow distribution behind the rear surface, where the reduction of the recirculation region leads to fast recovery of the streamwise velocity in a spanwise direction. \par

In the context of the control strategy being discussed, it is important to consider the forces acting on the surfaces of the cylinder. These forces can be separated into two components: pressure forces ($F_P$) and friction forces ($F_F$). The streamwise components of these force components contribute to the drag force, while the spanwise components contribute to the lift force. The pressure forces arise from the difference in pressure between the upstream and downstream sides of the cylinder, which explains the control strategy for Case 1, where the induced flow generated by the plasma actuators arranged downstream of the top-bottom surfaces results in a pressure gradient behind the rear surface and compensates the drag generated upstream of the cylinder. On the other hand, the friction force results from the interaction between the fluid and mainly the top and bottom surfaces of the square cylinder. This indicates that in Case 3, the control strategy impacts both the friction and pressure components of the forces applied on the cylinder, resulting in an effective control strategy in mitigating aerodynamic instabilities and improving overall flow control performance. The lift and drag forces can be defined as:

\begin{equation} \label{eqn:eq15}
F_D=F_{DP}+F_{DF}
, 
\end{equation}
\begin{equation} \label{eqn:eq16}
F_L=F_{LP}+F_{LF}
, 
\end{equation}

\noindent where $F_{DP}$ and $F_{DP}$ are the pressure and friction components of drag forces, and $F_{LP}$ and $F_{LF}$ are the pressure and friction components of the lift forces. \par

\begin{figure}
\centering 
\scalebox{0.5}{\includegraphics[angle=0, trim=0 0 0 0, width=0.9\textwidth]{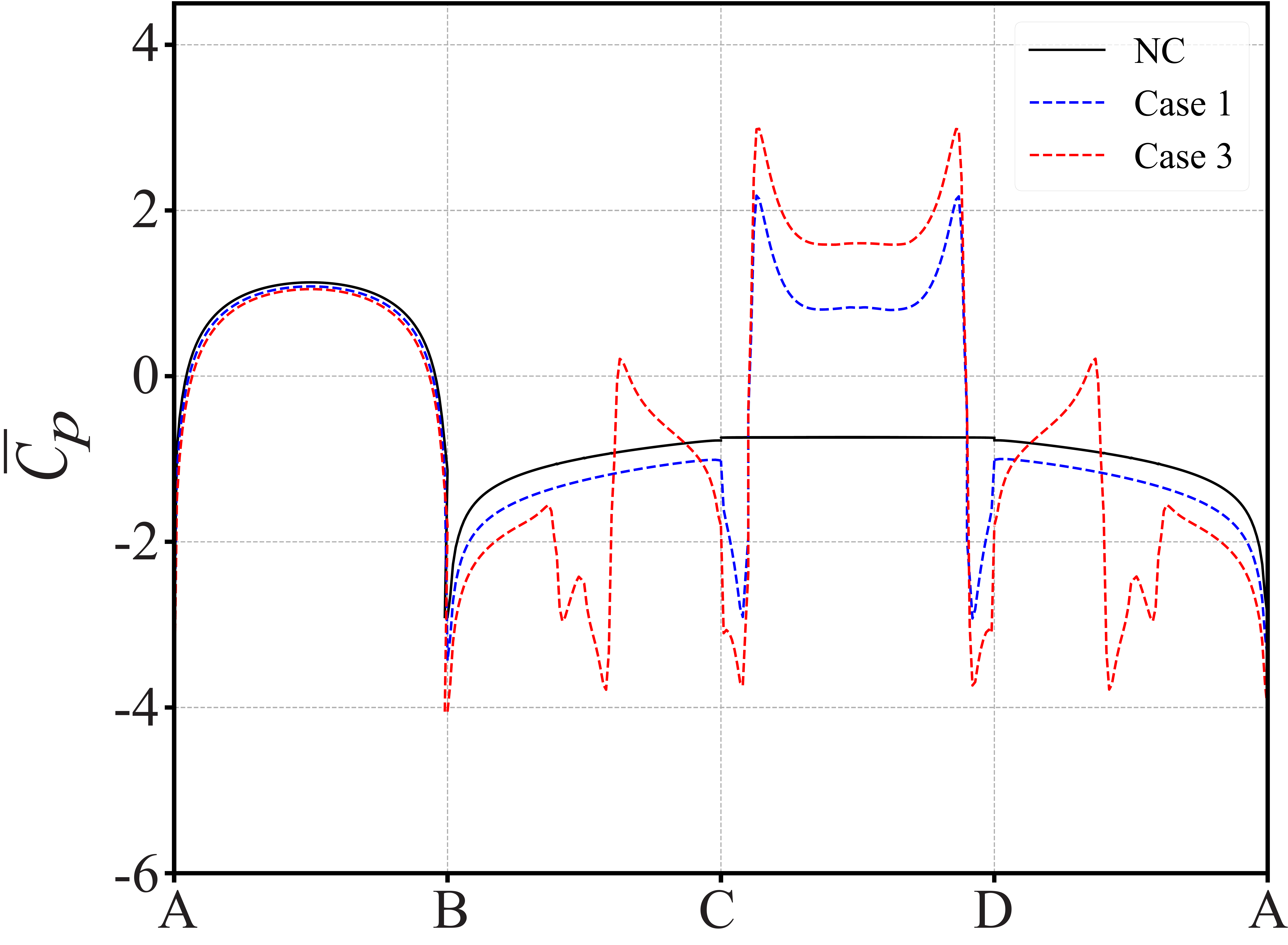}}
\caption[]{Pressure distribution on the cylinder surface at ${{Re}_D = 180}$.}
\label{fig:15-pressure}
\end{figure}

Figure~\ref{fig:15-pressure} provides insight into the pressure distribution on the surface of the cylinder, confirming that the control strategy for Case 1 primarily influences the rear surface of the cylinder, while for Case 3, it impacts the rear, top, and bottom surfaces of the cylinder.\par

\begin{figure}
\centering 
\includegraphics[angle=0, trim=0 0 0 0, width=0.9\textwidth]{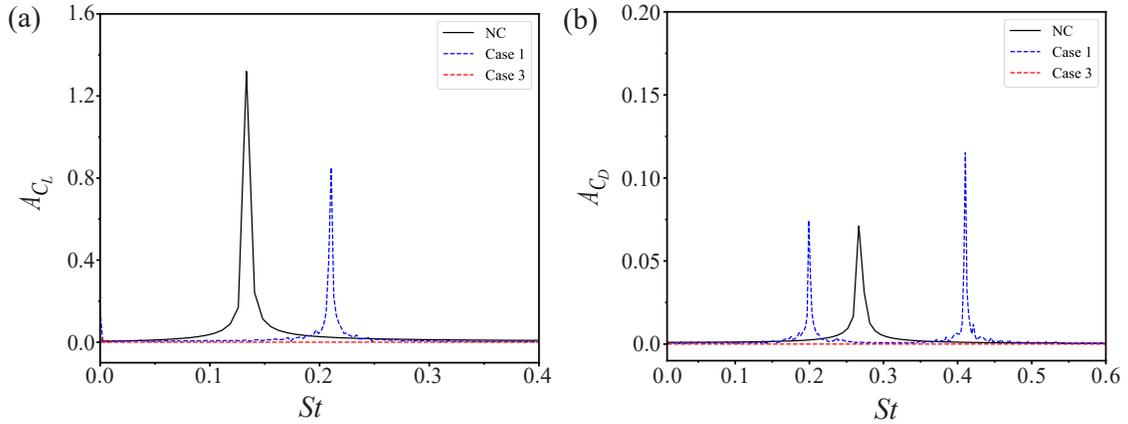}
\caption[]{FFT analysis of the drag and lift coefficients at ${Re}_D = 180$: (a) lift coefficient; (b) drag coefficient.}
\label{fig:16-fft}
\end{figure}

The results of the Fast Fourier Transform (FFT) analysis of the drag and lift coefficients for controlled cases are presented in Figure~\ref{fig:16-fft}. In the absence of control, the predominant frequencies coincide with the vortex shedding frequency in the wake. In Case 1, the shedding frequency of the lift coefficient is slightly elevated, while the oscillation amplitude is noticeably reduced. On the other hand, the oscillation amplitude and frequency of the drag coefficient are changed to two different values, indicating incomplete suppression of vortex shedding, with distinct primary frequency compositions. This might mainly be caused by the oscillatory behavior of the applied action as mentioned before. Conversely, in Case 3, there are no fluctuations in the lift and drag coefficients, implying a complete elimination of the vortex shedding. \par

\subsection{SHAP value analysis}\label{subsec: B3}

In this subsection, we employ SHAP values method proposed by Lundberg $et$ $al.$\cite{Lundberg2017} to determine the contribution of each feature (pressure probe in this study) to the prediction of the model. SHAP is based on the concept of the Shapley values\cite{Shapley1951, Roth1988} originated in cooperative game theory. The SHAP values provide a comprehensive and consistent method for attributing feature importance, enabling better model understanding, and interpretation in feature selection, and model comparison. For a given specific instance $x$, the SHAP value of each feature is denoted as $\phi_i(x)$, representing its fair contribution to the model's prediction. These values satisfy essential axioms: local accuracy ensures that the sum of SHAP values for all features equals the model's prediction for that instance $\phi_0(x) + \phi_1(x) + ... + \phi_n(x) = f(x)$, where n is the number of features, while missingness dictates that features irrelevant to the prediction have zero SHAP values, $\phi_i(x) = 0.$ Additionally, consistency ensures that differences in SHAP values between instances with a single feature variation correspond to the differences in their predictions $f(x) - f(x') = \phi_i(x) - \phi_i(x')$. More details can be found in the reference paper.\cite{Lundberg2017}



%

%





\begin{figure}
\centering 
\scalebox{1.1}{\includegraphics[angle=0, trim=0 0 0 0, width=0.9\textwidth]{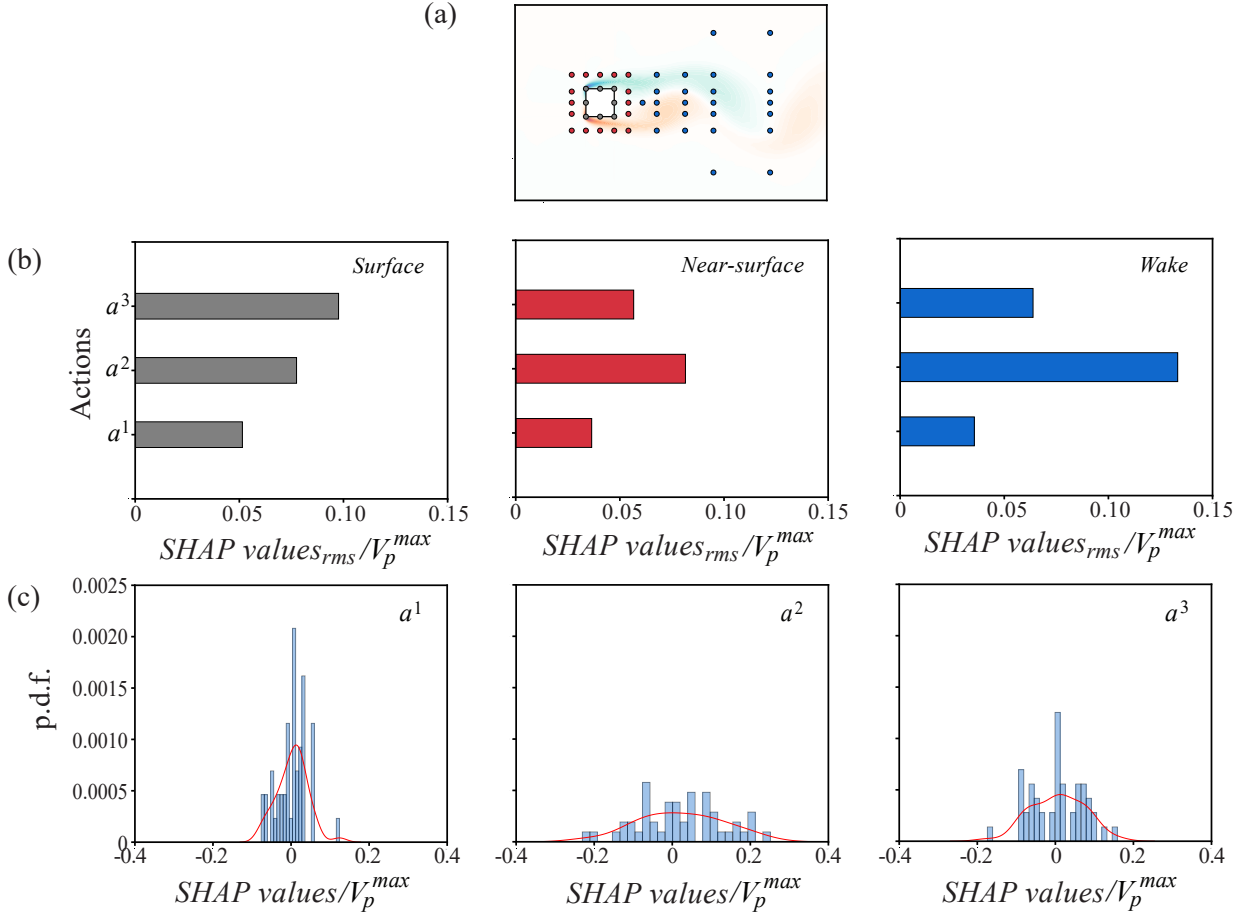}}
\caption[]{Analysis of the SHAP values for Case 3 at $Re_D=180$: (a) distribution of the pressure probes; (b) the rms SHAP values for representative areas of the probes; (c) the probability density function of the produced SHAP values.}
\label{fig:17SHAP}
\end{figure}

In this study, analysis of the trained model is conducted for the distributed pressure probes in the domain for Case 3 at $Re_D=180$, as shown in Figure~\ref{fig:17SHAP} (a). The probes that are colored in grey represent the surface region, including all the probes aligned on the surface of the cylinder. The near-surface region involves probes colored in red, and the probes with blue color represent the wake region. The results show the contribution values of the probes for each action taken by the agent at one instant. 
Results of the analysis are shown in Figure~\ref{fig:17SHAP} (b) and (c), where the rms of SHAP values for action 1 are shown to be the lowest with equal contributions from the wake and near-surface regions, and primary contribution from probes located at the cylinder's surface. This implies that action 1 is more likely to be fixed and unaffected by changes in flow parameters. Conversely, action 2 is highly dependent on the wake region, indicating that it is more likely to be regulated and has a significant contribution to suppressing vortex shedding in the wake. This suggests that action 2 is a crucial component of the active flow control strategy and requires careful optimization to achieve effective vortex shedding suppression. Regarding action 3, the analysis reveals a moderate contribution, with a high dependence on probes located on the surface of the cylinder. This implies that action 3 may have a more localized effect and could be used in conjunction with other control strategies to achieve enhanced vortex shedding suppression. Therefore, optimizing the contributions of all three actions would be essential for designing an effective flow control system. \par

Furthermore, by examining the probability density function (p.d.f.) distributions of SHAP values for each action, we observe that SHAP values near the zero are more prevalent for action 1, implying that each probe in the model tends to contribute equally. This suggests a low correlation between the instantaneous changes in flow parameters and action 1. Conversely, for action 2, the p.d.f. is widely distributed with a high variance of contributing values, indicating the sensitivity of this action. As for action 3, it exhibits a moderate contribution to the control strategy. 

\section{Conclusions}\label{sec:Conclusions}

We investigated the application of DRL in the context of AFC over a square cylinder, utilizing PPO as the learning algorithm, and incorporating a plasma actuator as the control mechanism. Due to the non-linearity associated with the proposed control configuration, which involves multiple plasma actuators operating within a specific range of applied AC voltage for each actuator, a neural network was trained to learn an optimal control policy. \par

The results demonstrated that PPO-based DRL is capable of effectively learning optimal control policies for AFC, leading to improved performance of the flow around the square cylinder. The trained agent was able to adapt its control actions in response to the varying flow conditions, resulting in remarkably reduced lift and drag coefficients, and consequently enhancing the aerodynamic stability of the cylinder. Initially, the efficiency of the designed flow control strategy was evaluated at $Re_D=100$ and subsequently applied to $Re_D=180$. The results demonstrated that the active flow control system significantly reduced the mean drag coefficient by 86\%, 93\%, and 97\% for each configuration at $Re_D=100$, and by 82\% and 99\% for cases 1 and 3 at $Re_D=180$. As the Reynolds number increased, it became increasingly challenging to eliminate vortex shedding using plasma actuators only located on the cylinder's rear surface. Nonetheless, the proposed control scheme succeeded in completely suppressing vortex shedding with an optimized configuration of the plasma actuators, indicating its potential for mitigating flow instabilities and reducing drag. \par

The control strategy involving the plasma actuator was shown to significantly impact the flow field, leading to improved flow control performance. Additionally, the SHAP values analysis of the three proposed control actions revealed distinct contributions to the flow control strategy. Action 1 had the lowest level of rms SHAP values, indicating a low correlation between instantaneous changes in flow parameters and this action. Action 2 had a widely distributed p.d.f. with a high variance of contributing values and high dependence on the wake region, suggesting its significant contribution to vortex shedding suppression in the wake. Action 3 showed a moderate contribution with high dependence on probes located on the surface of the cylinder, implying a more localized effect that could be used in conjunction with other control strategies. Optimizing the contributions of all three actions were found to be crucial for designing an effective flow control system.\par

The findings of this study contribute to the understanding of DRL-based AFC using PPO and plasma actuators, highlighting the potential of combining these techniques for achieving flow control. The results also suggest opportunities for further research, such as exploring different DRL algorithms, optimizing the control strategy of the plasma actuator, and investigating the applicability of the proposed approach in real-world experiments. \par

\begin{acknowledgments}
This work was supported by 'Human Resources Program in Energy Technology' of the Korea Institute of Energy Technology Evaluation and Planning (KETEP), granted financial resource from the Ministry of Trade, Industry \& Energy, Republic of Korea (no. 20214000000140). In addition, this work was supported by the National Research Foundation of Korea (NRF) grant funded by the Korea government (MSIP) (no. 2019R1I1A3A01058576). R.V. acknowledges the financial support from the ERC Grant No. "2021-CoG-101043998, DEEPCONTROL".
\end{acknowledgments}

\section*{Data Availability}
The data that supports the findings of this study are available within this article.

\appendix

\section*{Appendix}
\renewcommand{\thesubsection}{\Alph{subsection}}

\subsection{Open-source code}

You can access the source code for this study by clicking \href{https://fluids.pusan.ac.kr/fluids/65416/subview.do?enc=Zm5jdDF8QEB8JTJGYmJzJTJGZmx1aWRzJTJGMTY1MzQlMkYxMTUwMTY1JTJGYXJ0Y2xWaWV3LmRvJTNGYmJzT3BlbldyZFNlcSUzRCUyNmlzVmlld01pbmUlM0RmYWxzZSUyNnNyY2hDb2x1bW4lM0QlMjZwYWdlJTNEMSUyNnNyY2hXcmQlM0QlMjZyZ3NCZ25kZVN0ciUzRCUyNmJic0NsU2VxJTNEJTI2cGFzc3dvcmQlM0QlMjZyZ3NFbmRkZVN0ciUzRCUyNg%3D%3D}{\color{blue} \bf here}. The DRL code used in this study is built upon the open-source DRL code developed by Rabault $et$ $al.$\cite{Rabault2019_1} The DRL code follows the structure of a PPO agent implemented using the Tensorforce library. For the simulation environment, we have employed OpenFOAM-6 for the implementation.

\nocite{*}
\newpage
\bibliography{my-bib}

\end{document}